\documentclass[aps, prd, twocolumn, nofootinbib,]{revtex4-1}

\usepackage[top=1in, bottom=1in, left=1in, right=1in]{geometry}

\usepackage{amsmath}
\usepackage{amsfonts}
\usepackage{wasysym}
\usepackage{graphicx}
\usepackage{comment}
\usepackage{tensor}

\def\filetype{pdf}
\def\path{}

\allowdisplaybreaks[1]

\begin{document}

\title{Are gravitating magnetic monopoles stable?}
\author{Ben Kain}
\affiliation{Department of Physics, College of the Holy Cross, Worcester, Massachusetts 01610, USA}

\begin{abstract}
\noindent
The gravitating Julia-Zee dyon is a particle-like solution with both electric and magnetic charge.  It is found in the Einstein-Yang-Mills-Higgs system of $SU(2)$ with a scalar field in the adjoint representation coupled to gravity.  Within the magnetic ansatz this system is reduced from describing dyons to describing the gravitating 't Hooft-Polyakov magnetic monopole.  The stability of the well-known static gravitating magnetic monopole solutions with respect to perturbations within the magnetic ansatz---so-called magnetic perturbations---is well studied, but their stability with respect to perturbations outside the magnetic ansatz---so-called sphaleronic perturbations---is not.  I undertake a purely numerical study by adding sphaleronic perturbations to gravitating magnetic monopole solutions and then dynamically evolving the system.  For large perturbations I find that the system heads toward a dyon configuration, as expected.  For sufficiently small perturbations, however, the system oscillates about the magnetic ansatz in a manner consistent with oscillations about a stable equilibrium.
\end{abstract} 

\maketitle


\section{Introduction}

$SU(2)$, when spontaneously broken by a real triplet scalar field, has as a classical solution the Julia-Zee dyon \cite{Julia:1975ff}, a spherically symmetric particle-like solution with both electric and magnetic charge.  Within the magnetic ansatz, a physical constraint which sets the electric charge of the $U(1)$ subgroup to zero, the theory no longer describes dyons and has as a classical solution the 't Hooft-Polyakov monopole \cite{tHooft:1974kcl,Polyakov:1974ek}, a spherically symmetric particle-like solution with only magnetic charge.  When coupled to gravity, the system has regular and black hole static dyon solutions \cite{Brihaye:1998cm,Brihaye:1999nn} and, within the magnetic ansatz, regular and black hole static monopole solutions \cite{VanNieuwenhuizen:1975tc, Lee:1991vy, Ortiz:1991eu, Breitenlohner:1991aa, Breitenlohner:1994di}.

The magnetic ansatz, which plays a central role in this work, is self-consistent, in that an evolution that begins within the magnetic ansatz stays within the magnetic ansatz.  As I explain below, it is implemented by setting a certain group of fields to zero.  Thus, if an evolution begins with the relevant fields set to zero, these fields stay zero throughout the evolution.

The stability of the static gravitating magnetic monopole solutions has been studied in some detail, but only with respect to magnetic perturbations, which are perturbations within the magnetic ansatz, where there is little question that stability exists in a large area of parameter space \cite{Lee:1991qs, Aichelburg:1992st, Maeda:1993ap, Tachizawa:1994wn, Hollmann:1994fm, kain}.  This means that, in this area of parameter space, a dynamic evolution that begins with initial data within the magnetic ansatz will settle down to a static monopole configuration and not, say, disperse all matter fields to infinity \cite{kain}.

In addition to magnetic perturbations, there are sphaleronic perturbations, which are perturbations to the magnetic ansatz itself.  As far as I am aware, sphaleronic perturbations to the static monopole solutions have not yet been studied---presumably because it is very difficult to do so analytically (or semianalytically)---and, consequently, it is an open question whether gravitating magnetic monopoles are stable.  To avoid the difficulties in a (semi)analytical stability analysis, I undertake a purely numerical study by adding a sphaleronic perturbation to gravitating monopole solutions and then dynamically evolving the system.  Performing the necessary evolutions requires code that can dynamically evolve the full gravitating dyon system.  As far as I am aware, this is the first time the gravitating dyon has been dynamically solved.

For relatively large perturbations, the system appears to relax toward a dyon configuration.  As the size of the perturbation is made smaller, the electric charge in the system decreases and the end states of the evolutions move toward the monopole.  For sufficiently small perturbations, the electric charge density oscillates about zero in a manner suggestive of oscillations about a stable equilibrium.  This in turn is suggestive of the gravitating monopole being stable with respect to sphaleronic perturbations and, hence, of the static gravitating monopole solutions being stable with respect to both magnetic and sphaleronic perturbations.

The dynamic evolution of systems related to the gravitating dyon system studied here has a rich history.  Choptuik et al.\ \cite{Choptuik:1996yg, Choptuik:1999gh} dynamically evolved pure  $SU(2)$ (i.e.\ unbroken and without a scalar field) in their study of black hole critical phenomena.  This was further studied in the same system by a number of authors \cite{Bizon:2010mp, Rinne:2014kka, Maliborski:2017jyf}, as were tails and other topics \cite{Zenginoglu:2008wc, Purrer:2008my, Bizon:2010mp, Rinne:2013qc, Bizon:2014nla, Bizon:2016flx}.  Millward and Hirschmann \cite{Millward:2002pk} studied critical phenomena in $SU(2)$ with a scalar field in the fundamental representation.  Sakai \cite{Sakai:1995ds} was the first to dynamically evolve the gravitating monopole and was interested in what happens when the scalar field vacuum value is near its upper limit.  I recently evolved the monopole system in a study of type III critical phenomena and stability with respect to magnetic perturbations \cite{kain} and in a study of type II critical phenomena \cite{kain2}.  Finally, Gundlach, Baumgarte, and Hilditch made a related type II study in a system with a scalar field and an $SU(2)$ Yang-Mills field, but with only gravitational interactions \cite{Gundlach:2019wnk}.  With the important exception of the work of Rinne et al.\ in \cite{Rinne:2013qc, Maliborski:2017jyf}, all of these papers worked within the magnetic ansatz.  Thus, there has been limited dynamical study of $SU(2)$ outside the magnetic ansatz.

In the next section, I present the equations that describe the time-dependent gravitating Julia-Zee dyon and discuss gauge choices, the magnetic ansatz, and boundary conditions.  In Sec.\ \ref{sec:numerics} I discuss numerics.  In Sec.\ \ref{sec:results} I study the stability of gravitating monopoles with respect to sphaleronic perturbations.  I conclude in Sec.\ \ref{sec:conclusion}.


\section{Equations, gauges, the magnetic ansatz, and boundary conditions}

In this section I give the equations which describe the gravitating dyon system.  I gave many (though not all) of these equations in \cite{kain}, to which I refer the reader for additional information.  After presenting the equations, I discuss gauge choices for the matter sector, the magnetic ansatz, and boundary conditions.


\subsection{Metric equations}

My study of monopoles and dyons is restricted to spherical symmetry.  The general spherically symmetric metric in the Arnowitt-Deser-Misner (ADM) formalism \cite{AlcubierreBook, BaumgarteBook} is
\begin{align} \label{spherical metric}
ds^2 
&= - \left(\alpha^2 - a^2 \beta^2 \right)dt^2 + 2a^2 \beta dr dt + a^2 dr^2
\notag \\
&\qquad + B r^2 \left( d\theta^2 + \sin^2\theta d\phi^2 \right),
\end{align}
where the metric functions $\alpha$, $\beta$, $a$, and $B$ are functions of $t$ and $r$ only and I use units such that $c=1$ throughout.  These four functions obey the Einstein field equations,
\begin{equation}
G_{\mu\nu} = 8\pi G T_{\mu\nu},
\end{equation}
where $T_{\mu\nu}$ is the energy-momentum tensor.

The code I use to dynamically evolve the system uses radial-polar spacetime gauge.  This gauge has the benefit of simplifying equations by setting $B=1$ and $\beta = 0$.  $a$ and $\alpha$, the only metric functions to be solved for then, obey the constraint equations
\begin{equation} \label{radial-polar metric equations}
\begin{split}
\frac{a'}{a} &= 
4\pi G r a^2\rho 
- \frac{a^2 - 1}{2r}
\\
\frac{\alpha'}{\alpha} &=
4\pi G r  a^2 S\indices{^r_r} +\frac{a^2-1}{2r},
\end{split}
\end{equation}
which follow from the Einstein field equations.  In (\ref{radial-polar metric equations}), primes denote $r$ derivatives and $\rho$ and $S\indices{^r_r}$ come from the energy-momentum tensor and are given below.  Although the code I use only makes use of the metric functions $a$ and $\alpha$, in the following I give the general form of equations for completeness. 


\subsection{Matter equations}

The matter content of the 't Hooft-Polyakov monopole and the Julia-Zee dyon is an $SU(2)$ Yang-Mills-Higgs theory, with gauge field $A_\mu^a$ and real scalar field $\phi^a$ in the adjoint representation, where $a = 1,2,3$ is the gauge index (which can equivalently be placed up or down).  For $SU(2)$ the generators satisfy $[T_a,T_b] = i\epsilon_{abc} T_c$, where $\epsilon_{abc}$ is the completely antisymmetric symbol with $\epsilon_{123} = 1$.  In the adjoint representation I define the components of the generator matrices as $\left(T_a\right)_{bc} = -i\epsilon_{abc}$ with normalization $\text{Tr}(T_a T_b) = 2\delta_{ab}$, where $\text{Tr}$ here and below indicates a trace over generator matrices.  Defining
\begin{equation}
\phi \equiv T^a\phi^a , \quad
A_\mu \equiv T^a A_\mu^a, \quad
F_{\mu\nu} \equiv T^a F^a_{\mu\nu},
\end{equation}
where a sum over repeated gauge indices is implied and $F^a_{\mu\nu}$ is the field strength, the Yang-Mills-Higgs Lagrangian is
\begin{equation} \label{YMH Lagrangian}
\mathcal{L}_{YMH} = -\frac{1}{2} \text{Tr} 
\left[ \left(D_\mu \phi\right) \left(D^\mu \phi\right)
\right] - V +\mathcal{L}_{SU(2)},
\end{equation}
where
\begin{equation} \label{L SU(2)}
\mathcal{L}_{SU(2)}=
-\frac{1}{8g^2} \text{Tr} \left(F_{\mu\nu} F^{\mu\nu} \right),
\end{equation}
$g$ is the gauge coupling constant,
\begin{equation} \label{less standard covariant D}
\begin{split}
D_\mu \phi &= \nabla_\mu \phi - i \left[A_\mu, \phi\right]
\\
F_{\mu\nu} &= \nabla_\mu A_\nu - \nabla_\nu A_\mu - i \left[ A_\mu, A_\nu \right],
\end{split}
\end{equation}
and $V$ is the scalar potential, whose form I give below.  

Spherical symmetry constrains the fields.  The general spherically symmetric $SU(2)$ gauge field takes the form \cite{Witten:1976ck, Bartnik:1988am, Volkov:1998cc}
\begin{equation} \label{gauge field}
\begin{split}
A_t &= T^3 u_t
\\
A_r &= T^3 u_r
\\
A_\theta &= T^1 w_2 + T^2 w_1
\\
A_\phi &= \left(-T^1 w_1 + T^2 w_2 + T^3 \cot\theta \right)\sin\theta,
\end{split}
\end{equation}
where $u_t$, $u_r$, $w_1$, and $w_2$ parametrize the gauge field and are functions of $t$ and $r$ only, and the real triplet scalar field takes the form
\begin{equation}
\phi = \frac{\varphi}{\sqrt{2}} T^3,
\end{equation}
where $\varphi$ is a canonically normalized real scalar field and is a function of $t$ and $r$ only. The components of the spherically symmetric field strength can be found, for example, in \cite{Choptuik:1999gh, kain}.

Witten showed that spherical symmetry breaks $SU(2)$ down to $U(1)$ \cite{Witten:1976ck}.  This can be shown explicitly by writing the pure $SU(2)$ Lagrangian (\ref{L SU(2)}), with gauge field (\ref{gauge field}), as a Lagrangian for a complex scalar field gauged under $U(1)$:
\begin{align} \label{SU(2) as U(1) Lagrangian}
\mathcal{L}_{SU(2)} &= 
 -\frac{2}{g^2 B r^2} (D_\mu w)(D^\mu w)^*
\\
&\qquad- \frac{1}{2g^2B^2 r^4} (1 - |w|^2)^2
- \frac{1}{4g^2}  f_{\mu\nu} f^{\mu\nu},
\notag
\end{align}
where $w=w_1 + i w_2$,
\begin{equation} \label{U(1) defs}
\begin{split}
D_\mu w &= \nabla_\mu w - ia_\mu w
\\
a_\mu &= (u_t, u_r, 0, 0)
\\
f_{\mu\nu} &= \nabla_\mu a_\nu - \nabla_\nu a_\mu.
\end{split}
\end{equation}
I thus find that $w$ acts as the complex ``scalar" field gauged under $U(1)$, but with noncanonical kinetic terms and an atypical ``scalar" potential.  The $SU(2)$ Lagrangian (\ref{SU(2) as U(1) Lagrangian}) is clearly invariant under a $U(1)$ gauge transformation,
\begin{equation} \label{U(1) symmetry}
u_i \rightarrow u'_i = u_i - \nabla_i \tau, \qquad
w \rightarrow w' = w e^{-i\tau},
\end{equation}
where $i=t,r$ and $\tau$ is the gauge parameter.  Since the spherically symmetric kinetic term for the actual scalar field is
\begin{equation}
-\frac{1}{2}\text{Tr}\left[ (D_\mu \phi) (D^\mu \phi) \right] = 
-\partial_\mu \varphi \partial^\mu \varphi 
- \frac{2}{Br^2}|w|^2 \varphi^2,
\end{equation}
we find that with $\varphi$ invariant under the $U(1)$ transformation the complete theory has a $U(1)$ symmetry.  This symmetry will be made use of when I fix the gauge below. 

The scalar potential for the monopole and dyon is
\begin{equation}
V = \frac{\lambda}{4} \left( \varphi^2 - v^2 \right)^2,
\end{equation}
where $\lambda$ is the self-coupling constant and $v$ is the vacuum value of $\varphi$.  This scalar potential spontaneously breaks $SU(2)$ down to $U(1)$ giving rise to massive vector bosons and a massive scalar field with masses
\begin{equation} \label{vector scalar mass}
m_V = gv, \qquad m_S = \sqrt{2\lambda} \, v.
\end{equation}

I gave the equations of motion which follow from the Einstein-Yang-Mills-Higgs Lagrangian $\mathcal{L}_{EYMH} = \sqrt{-g} \mathcal{L}_{YMH}$ in \cite{kain}, which I repeat here.  For numerical purposes it is important to have the equations of motion in first-order form.  I thus define
\begin{align} \label{1st order var def}
\Phi &\equiv \varphi'&
\Pi &\equiv  
\frac{a B}{\alpha} \left(\dot{\varphi} - \beta\Phi \right)
\notag \\
Q_1 &\equiv w_1' + u_r w_2&
P_1 &\equiv   \frac{a}{\alpha} 
  \Bigl( \dot{w}_1 + u_t w_2 - \beta Q_1 \Bigr)
\notag \\
Q_2 &\equiv w_2' - u_r w_1&
P_2 &\equiv \frac{a}{\alpha } 
\Bigl( \dot{w}_2 - u_t w_1 - \beta Q_2 \Bigr)
\notag \\
Y &\equiv \frac{B r^2}{2 \alpha a} \left(\dot{u}_r - u'_t \right).
\end{align}
I list the equations of motion grouped into families, using a dot to denote $t$ derivatives.  First $\varphi$, $\Phi$, and $\Pi$:
\begin{align}
\dot{\varphi} &= \frac{\alpha}{aB} \Pi + \beta \Phi 
\notag\\
\dot{\Phi} &=
\partial_r \left( \frac{\alpha}{a B} \Pi + \beta \Phi \right) 
\notag\\
\dot{\Pi} &= 
\frac{1}{r^2} \partial_r \left( \frac{\alpha B r^2}{a} \Phi + r^2 \beta \Pi \right)
- \alpha a B  
\frac{\partial V}{\partial \varphi}
\notag\\
&\qquad - \frac{2\alpha a}{r^2} (w_1^2 + w_2^2) \varphi,
\label{varphi evo eqs}
\end{align}
then $w_1$, $Q_1$, and $P_1$:
\begin{align}
\dot{w}_1 &= \frac{\alpha}{a} P_1 - u_t w_2 + \beta Q_1 
\notag \\
\dot{Q}_1 &=
\partial_r  \left(\frac{\alpha}{a}P_1 + \beta Q_1 \right)
- u_t Q_2
 + u_r \left(\frac{\alpha}{a}P_2 + \beta Q_2 \right) 
\notag \\
&\qquad
+ w_2\frac{2\alpha a}{Br^2} Y
\notag \\
\dot{P}_1 &= \partial_r \left( \frac{\alpha}{a} Q_1 + \beta P_1 \right) - P_2 (u_t - \beta u_r )
 + \frac{\alpha}{a} u_r Q_2
\notag \\
&\qquad
 + \frac{\alpha a}{B r^2}w_1 (1 -w_1^2 - w_2^2)
- g^2 \alpha a w_1 \varphi^2,
\end{align}
and $w_2$, $Q_2$, and $P_2$:
\begin{align}
\dot{w}_2 &= \frac{\alpha}{a} P_2 + u_t w_1 + \beta Q_2
\notag \\
\dot{Q}_2 &=
\partial_r  \left(\frac{\alpha}{a}P_2 + \beta Q_2 \right)
+ u_t Q_1
- u_r \left(\frac{\alpha}{a} P_1 + \beta Q_1 \right)
\notag \\
&\qquad
- w_1 \frac{2\alpha a}{Br^2} Y 
\notag \\
\dot{P}_2 &= \partial_r \left( \frac{\alpha}{a} Q_2 + \beta P_2 \right) + P_1 (u_t - \beta u_r ) - \frac{\alpha}{a} u_r Q_1 
\notag \\
&\qquad
+ \frac{\alpha a}{B r^2}w_2 (1 -w_1^2 - w_2^2)
- g^2 \alpha a w_2 \varphi^2,
\label{w2 evo eqs}
\end{align}
and finally $u_r$ and $Y$:
\begin{align}
\dot{u}_r &= \frac{2\alpha a}{Br^2}Y + u'_t 
\label{ur q evo eqs}\\
\dot{Y} &= \frac{\alpha}{a} \left(w_1 Q_2 - w_2 Q_1\right) + \beta \left(w_1 P_2 - w_2 P_1 \right).
\notag
\end{align}
Note that I do not have an evolution equation for $u_t$, which I will handle when fixing the gauge.  There exists one final equation, which is the Gauss constraint:
\begin{equation} \label{Gauss constraint}
Y' = w_1 P_2  - w_2 P_1.
\end{equation}
We shall see below that $Y$ is proportional to the total electric charge inside a sphere of radius $r$ and thus $Y' = \partial Y / \partial r$ is proportional to the radial electric charge density.

In \cite{kain} I gave the energy-momentum tensor that follows from the Lagrangian in (\ref{YMH Lagrangian}), including each of its nonvanishing components and a number of commonly used  matter functions which follow from it.  Here I repeat only the matter functions used in (\ref{radial-polar metric equations}):
\begin{widetext}
\begin{equation} \label{energy-momentum functions}
\begin{split} 
\rho &=
\frac{1}{2a^2} \left( \Phi^2 + \frac{\Pi^2}{B^2} \right)
+ \frac{(w_1^2 + w_2^2)\varphi^2}{Br^2} + V
+ \frac{(1 - w_1^2 - w_2^2)^2 }{2g^2 B^2 r^4}
+ \frac{Q_1^2 + Q_2^2 + P_1^2 + P_2^2 }{g^2a^2 Br^2}
+ \frac{2Y^2}{g^2 B^2 r^4}
\\
S\indices{^r_r} &=
\frac{1}{2a^2} \left( \Phi^2 + \frac{\Pi^2}{B^2} \right)
- \frac{(w_1^2 + w_2^2)\varphi^2}{Br^2} - V
- \frac{(1 - w_1^2 - w_2^2)^2 }{2g^2 B^2 r^4}
+ \frac{Q_1^2 + Q_2^2 + P_1^2 + P_2^2 }{g^2a^2 Br^2}
- \frac{2Y^2}{g^2 B^2 r^4} .
\end{split}
\end{equation}
\end{widetext}


\subsection{Electric charge and mass}

The electric charge is found with the help of the conserved electric current, $j^\mu$, which follows from the inhomogeneous Maxwell equation,
\begin{equation}
\nabla_\mu f^{\mu\nu} = g j^\nu,
\end{equation}
where the factor of $g$ is included because of my convention for the $U(1)$ gauge field in (\ref{SU(2) as U(1) Lagrangian}) and (\ref{U(1) defs}).  That $\nabla_\mu j^\mu = 0$, and hence that $j^\mu$ is conserved, follows immediately from $f^{\mu\nu}$ being antisymmetric.  The components of the current work out to be
\begin{equation}
j^t = \frac{2 Y'}{g \alpha a B r^2}, \qquad
j^r = -\frac{2 \dot{Y}}{g \alpha a B r^2}.
\end{equation}
The total charge enclosed in a sphere of radius $r$ is given by \cite{Petryk:2006pg}
\begin{align}
q(t,r) &= 
\int \sqrt{\gamma} (-n_\mu j^\mu) dr d\theta d\phi
=  \frac{8\pi}{g} \int_0^r Y' dr
\notag \\
&=  \frac{8\pi}{g} Y(t,r),
\end{align}
where $n_\mu = (-\alpha,0,0,0)$ is the time-like unit vector normal to the spatial slices, $-n_\mu j^\mu$ is the electric charge density, and $\gamma = a^2 B^2 r^4 \sin^2\theta$ is the determinant of the spatial metric.  I explain below that the finiteness of the energy density at the origin requires $Y(t,0) = 0$, which allows for the evaluation of the limits above.  The total charge in the system, $q_\infty \equiv q(t,\infty)$, is a conserved quantity.  As promised, $Y$ is proportional to the total electric charge inside a sphere of radius $r$.  I note that $q$ is related to the radial component of the electric field, $E^r = - f^{r\mu}n_\mu/g
=  q/(4\pi a B r^2)$, where the factor of $g$ in the definition for $E^r$ again follows from my convention for the $U(1)$ gauge field in (\ref{SU(2) as U(1) Lagrangian}) and (\ref{U(1) defs}).

A convenient form for the mass function can be motivated by looking at the static solution in the large $r$ limit.  Defining for convenience the function $\mathcal{M}$ as
\begin{equation}
\frac{1}{g_{rr}} = \frac{1}{a^2} \equiv 1 - \frac{2G \mathcal{M}}{r},
\end{equation}
from which $G\mathcal{M} = (r/2)(1-1/a^2)$, I have
\begin{equation}
 \mathcal{M}' = 4\pi  r^2 \rho,
\end{equation}
where I used the $a'$ equation in (\ref{radial-polar metric equations}).  I explain below that the outer boundary conditions at $r= \infty$ are $\varphi = \pm v$ and $w_1 = w_2 = 0$, for which, in the static limit, the energy density,  $\rho$ in (\ref{energy-momentum functions}), reduces significantly and
\begin{equation}
\mathcal{M}' = \frac{2\pi}{g^2 r^2} + \frac{q^2}{8\pi r^2}.
\end{equation}
In the large $r$ limit $q\rightarrow q_\infty$ and is constant, allowing the above equation to be easily integrated and we have 
\begin{equation}
\frac{1}{g_{rr}} = \frac{1}{a^2} = 1 - \frac{2 G M}{r} + \frac{4\pi G(1/g)^2}{r^2} + \frac{G q_\infty^2/4\pi}{r^2},
\end{equation} 
where $M$ is the ADM mass. This is the Reissner-Nordstr\"om solution with unit magnetic charge (in units of $g$) and electric charge $q_\infty$.  This solution motivates defining the mass function as
\begin{equation} \label{mass function}
m(t,r) = \frac{r}{2G} \left[ 1 - \frac{1}{a^2(t,r)} + \frac{4\pi G}{g^2 r^2} + \frac{G q^2(t,r)}{4\pi r^2} \right],
\end{equation}
whose asymptotic value gives the ADM mass $M = m(t,\infty)$.


\subsection{Matter gauges}

The matter sector obeys the gauge transformation (\ref{U(1) symmetry}) and it will be useful to fix this gauge.  In this subsection I comment on a few gauge choices.  One choice is temporal gauge, which fixes $u_t = 0$ and immediately solves the problem that there is no evolution equation for $u_t$.  In some gauges, static solutions---in which gauge-invariant fields, such as $|w|=\sqrt{w_1^2 + w_2^2}$ and $Y$, are time independent---have gauge-dependent fields, such as $u_r$, $w_1$, and $w_2$, which retain a time dependence.  Temporal gauge is perhaps the easiest gauge in which to see that this is the case.  $Y$, being proportional to the total electric charge inside a sphere of radius $r$, must be time independent and nonzero for a static dyon.  A look at its definition in (\ref{1st order var def}) shows that with $u_t = 0$, $\dot{u}_r$ must be nonzero.

Another possibility is radial gauge, which fixes $u_r = 0$ and the $u_r$ evolution equation in (\ref{ur q evo eqs}) reduces to an ODE for $u_t$.  Rinne et al.\ used radial gauge in their dynamical study of pure $SU(2)$ \cite{Rinne:2013qc, Maliborski:2017jyf}.  Radial gauge is the best choice for finding static solutions directly, since, for static solutions, all fields are time independent and one can additionally fix $w_2 = 0$.

The final gauge I mention is Lorenz gauge, which introduces an evolution equation for $u_t$ through the Lorenz gauge condition, $\nabla_\mu a^\mu = 0$.  As with temporal gauge, some gauge-dependent fields in Lorenz gauge retain a time dependence for static solutions.  I use Lorenz gauge in this work because, for the numerical scheme I am using, I found Lorenz gauge to be the most stable.  Introducing the auxiliary field
\begin{equation} \label{Omega def}
\Omega \equiv \frac{a B}{\alpha}(u_t - \beta u_r),
\end{equation}
the Lorenz gauge condition can be written as
\begin{equation} \label{Lorenz eqs}
u_t = \frac{\alpha}{aB} \Omega + \beta u_r, 
\quad
\dot{\Omega} = \frac{1}{r^2} 
\partial_r \left[ r^2 \left( \frac{\alpha B}{a} u_r + \beta \Omega\right)\right].
\end{equation}


\subsection{Magnetic ansatz}
\label{sec:magnetic ansatz}

The magnetic ansatz is a physical constraint on the theory (and not a gauge choice) which sets the electric charge of the Abelian subgroup to zero and reduces the dyon to the monopole.  I take its definition to be
\begin{equation}
Y'(t,r) = 0,
\end{equation}
since $Y'$ is proportional to the electric charge density.\footnote{It is easy to see that if $Y'(t,r)=0$, then physically it must also be that $Y(t,r)=0$, since $Y(t,r)$ is proportional to the total electric charge inside a sphere of radius $r$. $Y(t,r) = 0$  is a common way of expressing the magnetic ansatz.}  Once the magnetic ansatz is made, convenient gauge choices (see, for example, \cite{Choptuik:1999gh, kain} for details) set $u_t = u_r = w_2 = 0$ and the only nonvanishing matter fields are $\varphi$ and $w_1$.  

The magnetic ansatz is self-consistent, in that an evolution that begins with initial data within the magnetic ansatz remains within the magnetic ansatz.  That this is so is a big reason why nearly all dynamical gravitational studies of $SU(2)$ have been done within the magnetic ansatz (the only exceptions I am aware of are \cite{Rinne:2013qc,  Maliborski:2017jyf}).  To be specific, an evolution with initial data that has $Y' = u_t = u_r = w_2 = 0$ everywhere, keeps $Y' = u_t = u_r = w_2 = 0$ everywhere.  An immediate consequence is that an evolution that begins with the gravitating monopole, stays with the gravitating monopole.

It is an open question whether the magnetic ansatz is stable in the gravitating monopole system and hence whether gravitating monopoles are stable.  I study this issue numerically in Sec.\ \ref{sec:results}.


\subsection{Boundary conditions}
\label{sec:boundary conditions}

To solve the system of equations I need boundary conditions for many of the variables.  Boundary conditions include both conditions at the boundary of space and the boundary of the computational domain.  I list a number of boundary conditions in this subsection and discuss the outer boundary of the computational domain in the next section.

The inner boundary condition for $a$ is $a(t,0) = 1$, which is the flat space value $a$ has when inside a spherically symmetric matter distribution and follows from finiteness of the top equation in (\ref{radial-polar metric equations}).  As can be seen from the bottom equation in (\ref{radial-polar metric equations}), any solution for $\alpha$ can be scaled by a constant and still be a solution.  I set $\alpha(t,r)=1/a(t,r)$ at large $r$, a choice motivated by the spacetime being asymptotically Reissner-N\"ordstrom.  I take the parity of $a$ and $\alpha$ to be even near the origin.

Some boundary conditions for matter functions follow from the energy density, $\rho$ in (\ref{energy-momentum functions}), being finite at the origin and $r^2\rho$ vanishing as $r\rightarrow \infty$ so that the total integrated energy is finite.  At the inner boundary I have $\varphi = O(r)$, $|w|^2=w_1^2 + w_2^2 = 1 + O(r^2)$, and  $Y = O(r^2)$.  Additional inner boundary conditions can be found by solving the equations of motion after expanding them around the origin.  I find
\begin{equation}
w_1 = \cos\theta_w(t) + O(r^2), \quad
w_2 = \sin\theta_w(t) + O(r^2),
\end{equation}
where I have introduced the angle $\theta_w$ as a parametrization of $w_1^2+w_2^2 = 1 + O(r^2)$, and
\begin{equation}
u_t = \dot{\theta}_w(t) + O(r^2), \qquad
u_r = O(r).
\end{equation}
I note in particular that the equation for $u_t$ is the solution to the Gauss constraint in (\ref{Gauss constraint}).  It is easy to see that the equation for $u_t$ may also be written as $u_t = -\dot{w}_1/w_2 + O(r^2)$ and $u_t = \dot{w}_2/w_1 + O(r^2)$.  These two forms are precisely what is needed for the $(P_1^2 + P_2^2)/r^2$ term in the energy density to be finite at the origin.  At $r= \infty$ I have $\varphi = \pm v$ and $w_1= w_2 =0$.  I take the parity of the matter fields to be $\Phi$, $w_1$, $w_2$, $P_1$, $P_2$, $u_t$, and $\Omega$ are even and $\varphi$, $\Pi$, $Q_1$, $Q_2$, $u_r$, and $Y$ are odd near the origin.


\section{Numerics}
\label{sec:numerics}

In this section I describe numerical aspects, including the code I use to dynamically evolve the system of equations listed in the previous section.  As mentioned there, I evolve the system in radial-polar spacetime gauge, which fixes $B= 1$ and $\beta = 0$.  The constraint equations in (\ref{radial-polar metric equations}) determine the metric functions $a$ and $\alpha$ on a given time slice and I solve them using second-order Runge-Kutta.  The evolution equations in (\ref{varphi evo eqs})--(\ref{ur q evo eqs}) and the Lorenz gauge condition equations in (\ref{Lorenz eqs}) determine the matter fields $\varphi$, $\Phi$, $\Pi$, $w_1$, $Q_1$, $P_1$, $w_2$, $Q_2,$ $P_2$, $u_t$, $u_r$, $Y$, and $\Omega$ and I solve them using the method of lines and third-order Runge-Kutta.  I note in particular that I solve for $Y$ using its evolution equation in (\ref{ur q evo eqs}) instead of the Gauss constraint in (\ref{Gauss constraint}) because I found this to be more stable.   I use centered sixth-order finite differencing for spatial derivatives.  In solving the evolution equations I include fourth-order Kreiss-Oliger dissipation \cite{AlcubierreBook} to help with stability.  Inner boundary conditions at the origin are as given in Sec.\ \ref{sec:boundary conditions}. 

Since the outer boundary of the computational domain does not extend to $r=\infty$ I need outer boundary conditions for the matter fields that allow them to exit the computational domain.  I use standard outgoing wave conditions with $\varphi$ modeled as a spherical wave and $w_1$ and $w_2$ modeled as one-dimensional waves, just as in \cite{kain}.  Additionally, I model $u_r$ and $\Omega$ as spherical waves.  $u_t$ and $Y$ do not need outer boundary conditions since $u_t$ is given by an algebraic equation in Lorenz gauge and the evolution equation for $Y$ does not contain spatial derivatives and can be integrated right up to the outer boundary.

In any numerical study it is best to use dimensionless variables.  In the literature there exist two common mass scales used for constructing dimensionless quantities:\ $m_P$ and $v$, where $m_P = 1/\sqrt{G}$ is the Planck mass and $v$ is the vacuum value of the scalar field.  As in \cite{kain}, I use $m_P$ and define $m_G \equiv m_P/\sqrt{4\pi}$ (where the $\sqrt{4\pi}$ is included for convenience) and the dimensionless quantities
\begin{gather}
\bar{r} \equiv (gm_G)r, \quad
\bar{t} \equiv (gm_G)t,
\notag \\
\bar{v} \equiv v/m_G, \quad
\bar{\lambda} \equiv \lambda/g^2,
\\
\bar{\varphi} \equiv \varphi/m_G, \quad
\bar{u}_t \equiv u_t/gm_G, \quad 
\bar{u}_r \equiv u_r/gm_G,
\notag
\end{gather}
along with $\bar{m} \equiv (gm_G/m_P^2)m$ and $\overline{\Omega} \equiv \Omega/gm_G$.  I note that $w_1$, $w_2$, and $Y$ are already dimensionless and $\bar{v} = m_V/gm_G$ and $\bar{\lambda} = (m_S/\sqrt{2}m_V)^2$, where $m_V$ and $m_S$ are the vector and scalar masses in (\ref{vector scalar mass}).  The results presented in the next section will be the radial energy and radial electric charge densities.  For future convenience, then, the dimensionless quantities in terms of the dimensionful quantities are
\begin{equation}
\bar{r}^2 \bar{\rho}
=4\pi r^2 \rho / m_P^2 , 
\qquad
Y' = \frac{\partial Y}{\partial \bar{r}} = \frac{1}{4\sqrt{\pi} m_P} \frac{\partial q}{\partial r},
\end{equation}
where $\bar{\rho} \equiv \rho/g^2 m_G^4$ is the dimensionless energy density and a prime now denotes a derivative with respect to $\bar{r}$ instead of $r$.

The code I use is second-order accurate and I have confirmed second order convergence.  In Lorenz gauge it is surprisingly stable.  I have not found any indications of instability using a uniform computational grid and a grid-point spacing of  $\Delta \bar{r} = 0.06$, or even larger, and a time step of $\Delta{\bar{t}}/\Delta{\bar{r}} = 0.5$, including for very long runs.  Further, there is no discernible  difference between results using  $\Delta \bar{r} = 0.06$ and a smaller grid-point spacing.  By using the relatively large grid-point spacing $\Delta \bar{r} = 0.06$, I can also use a large value for $\bar{r}_\text{max}$, the position of the outer boundary, and still have run times that are not impractical.  Any numerical scheme that allows fields to exit the computational domain will have (artificial) reflections due to fields not perfectly exiting.  By pushing the outer boundary far enough out, these reflections will take so long to return that they cannot influence what happens near the origin.  For the results presented in the next section I use $\Delta\bar{r} = 0.06$,   $\Delta{\bar{t}}/\Delta{\bar{r}} = 0.5$, $\bar{r}_\text{max} = 5000$, and evolve the system to $\bar{t} = 10\,000$.


\section{Sphaleronic stability of gravitating monopoles}
\label{sec:results}

\begin{figure*}
\centering
\includegraphics[width=6.5in]{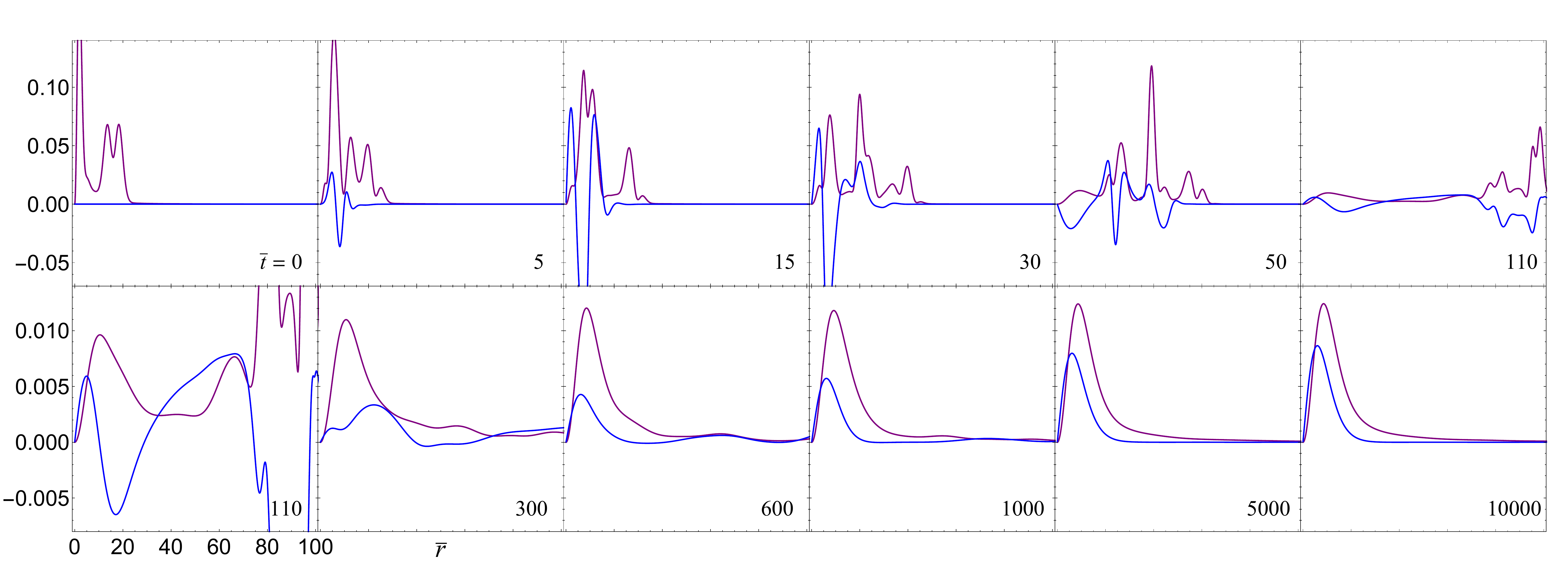}
\caption{A time evolution of the radial energy density, $\bar{r}^2\bar{\rho}$ (purple), and $Y' = \partial Y/\partial \bar{r}$ (blue), which is proportional to the radial electric charge density, as a function of $\bar{r}$ for $\bar{v} = 0.2$, $\lambda = 0$, initial data (\ref{initial data}) with  $\bar{s}_\varphi = 10$, $\bar{r}_{1} = 2$, and $\bar{s}_{1} = 5$, and sphaleronic perturbation (\ref{w2 IC}) with $\bar{r}_{2} = 15$, $\bar{s}_2 = 4$, and $f_2$ = 1.  Although only plotted out to $\bar{r} = 100$, the outer boundary of the computational domain extends to $\bar{r} = 5000$.  The perturbation is large and the evolution appears to head toward a dyon configuration and maintain a large nonzero value for $Y'$, which breaks the magnetic ansatz.  Starting in the bottom row, I change the vertical scale to better see the solutions.  The value of $\bar{t}$ is given in the corner of each frame.}
\label{fig:evo 1}
\end{figure*}

In this section I study the stability of gravitating monopoles with respect to sphaleronic perturbations, i.e.\ perturbations to the magnetic ansatz.  I do so by taking initial data within the magnetic ansatz, and thus initial data for a gravitating monopole, and adding to it a magnetic ansatz-breaking perturbation.  I then dynamically evolve the system.  My focus will primarily be on the quantity $Y'$.  This is because $Y'$ is gauge invariant and $Y' = 0$ defines the magnetic ansatz.\footnote{One can just as easily use $Y$ instead of $Y'$ and obtain the same results found below.}  

The parameters of the system are $\bar{v}$ and $\bar{\lambda}$.  In the following I restrict attention to $\bar{v} = 0.2$ and $\bar{\lambda} = 0$.  For these values, there exists a unique regular static monopole solution \cite{Breitenlohner:1991aa}.\footnote{I am referring to the fundamental solution, in which the gauge field $w_1$ only equals zero at $r=\infty$, and not to excited solutions \cite{Breitenlohner:1991aa} which are expected to be unstable.}  This means that all initial data with $\bar{v} = 0.2$, $\bar{\lambda} = 0$, and without a sphaleronic perturbation evolves to the same static monopole solution (as long a black hole does not form) \cite{kain}.  Further, it means that all sphaleronic perturbations are perturbing the same static monopole solution.  

I explained in Sec.\ \ref{sec:magnetic ansatz} that the magnetic ansatz can be thought of as $Y' = 0$, along with $u_t = u_r = w_2 = 0$ (the latter set of conditions being gauge dependent).  Thus, in constructing initial data, I begin with $Y' = u_t = u_r = w_2 = 0$ and then add a nonzero value to one of these fields.  For reasons having to do with constructing initial data, I only consider nonzero values for $w_2$ and $u_r$.  I present results for a $w_2$ perturbation to generic magnetic initial data and a $u_r$-perturbation to the static gravitating monopole solution.  I have studied evolutions for various initial data and found the results that follow to be typical.
\begin{figure*}
\centering
\includegraphics[width=6.5in]{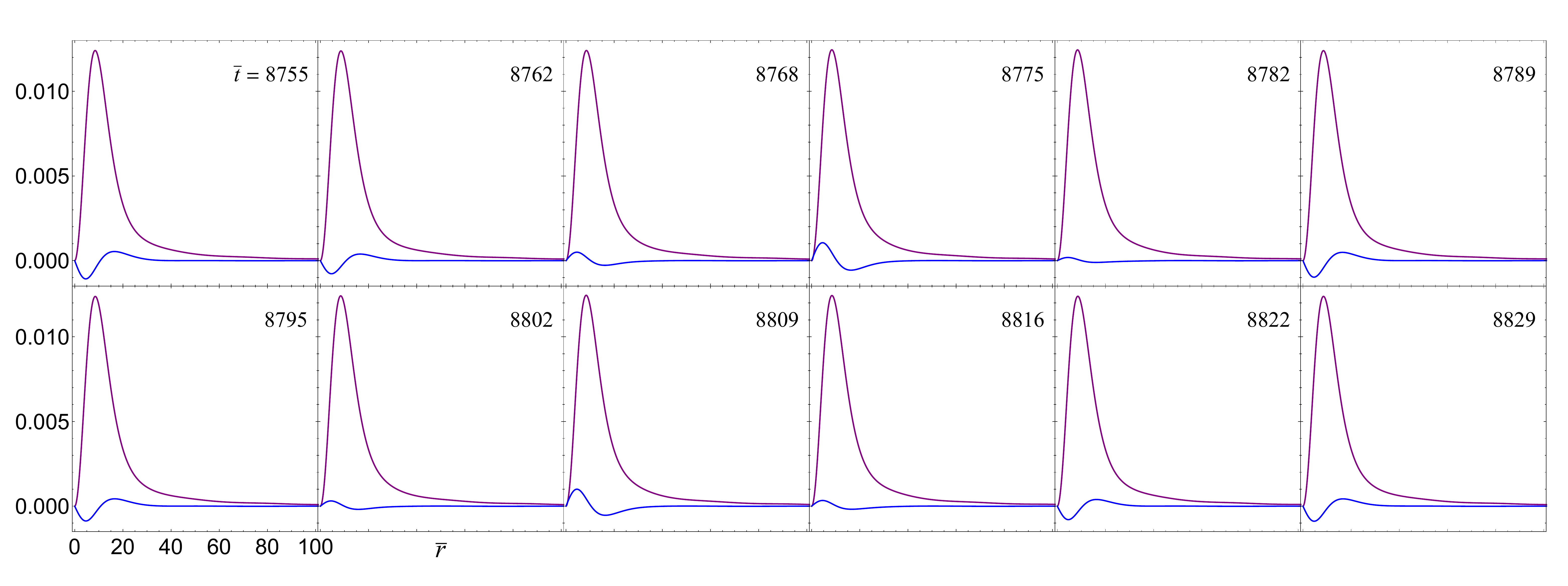}
\caption{A time evolution of the radial energy density, $\bar{r}^2\bar{\rho}$ (purple), and $Y' = \partial Y/\partial \bar{r}$ (blue), which is proportional to the radial electric charge density, as a function of $\bar{r}$ with the same initial data given in the caption of Fig.\ \ref{fig:evo 1}, except with the sphaleronic perturbation strength $f_2 = 0.02$, making this a small perturbation.  The value of $\bar{t}$ is given in the corner of each frame.  I do not show the beginning of the evolution because it looks similar to that shown in Fig.\ \ref{fig:evo 1}.  I show instead late times where we can see $Y'$ oscillating.}
\label{fig:evo 2}
\end{figure*}

For the $w_2$ perturbation, I use for the magnetic part of the initial data \cite{kain}
\begin{align}
\varphi(0,r) &= v \tanh \left( r/s_\varphi \right)
\notag \\
w_1(0,r) &= 
\frac{1}{2}
\Biggl\{1 + 
\biggl[1 + a_1 \left(1 + \frac{b_1 r}{s_1}\right) 
e^{-2(r/s_1)^2} \biggr]
\notag \\
&\qquad\qquad\qquad \times
\tanh \left( \frac{r_1-r}{s_1}\right)
\Biggr\}, \label{initial data}
\end{align}
along with $\dot{\varphi}(0,r) = \dot{w}_1(0,r) = 0$.  The parameters $r_1$ and $s_1$ give the center and spread of the $w_1$ pulse and the parameters $a_1$ and $b_1$ are chosen such that the boundary conditions for $w_1$ are satisfied at the origin and are given by
$a_1 = \coth(r_1/s_1)- 1$ and $b_1 = \coth(r_1/s_1)+1$.  The sphaleronic $w_2$ perturbation is a Gaussian:
\begin{equation} \label{w2 IC}
w_2(0,r) = f_2 
(r/r_2 )^2
e^{-(r-r_2)^2/s_2^2},
\end{equation}
along with $\dot{w}_2(0,r) = 0$.  The parameters $r_2$ and $s_2$ give the center and spread of the perturbation and $f_2$ can be thought of as its strength.

In Fig.\ \ref{fig:evo 1} I show a typical evolution for a large perturbation.  The main purpose of this figure is to give an impression of what an evolution looks like.  I plot the radial energy density,  $\bar{r}^2\bar{\rho}$ (purple), and $Y'=\partial Y/\partial\bar{r}$ (blue), which is proportional to the radial electric charge density.  One can thus see how energy and electric charge distribute themselves over the course of an evolution.  Both the energy and electric charge appear to maintain a localized configuration at late times and thus the system appears to settle toward a gravitating dyon.  The expectation for a large perturbation is that the system is pushed far from the monopole and stays far from the monopole.  That the evolution in Fig.\ \ref{fig:evo 1} appears to maintain a large nonzero electric charge density is consistent with this.

I show a typical evolution when the perturbation is small in Fig.\ \ref{fig:evo 2}, where again the purple curve is the radial energy density and the blue curve is $Y'$, which is proportional to the radial electric charge density.  The beginning of the evolution is similar to Fig.\ \ref{fig:evo 1} and is not shown.  I focus instead on late times where we can see $Y'$ oscillating.  Physically, it would appear that shells of positive and negative charge trade places as they oscillate closer and then farther from the center of the system. 

\begin{figure*}
\centering
\includegraphics[width=6.5in]{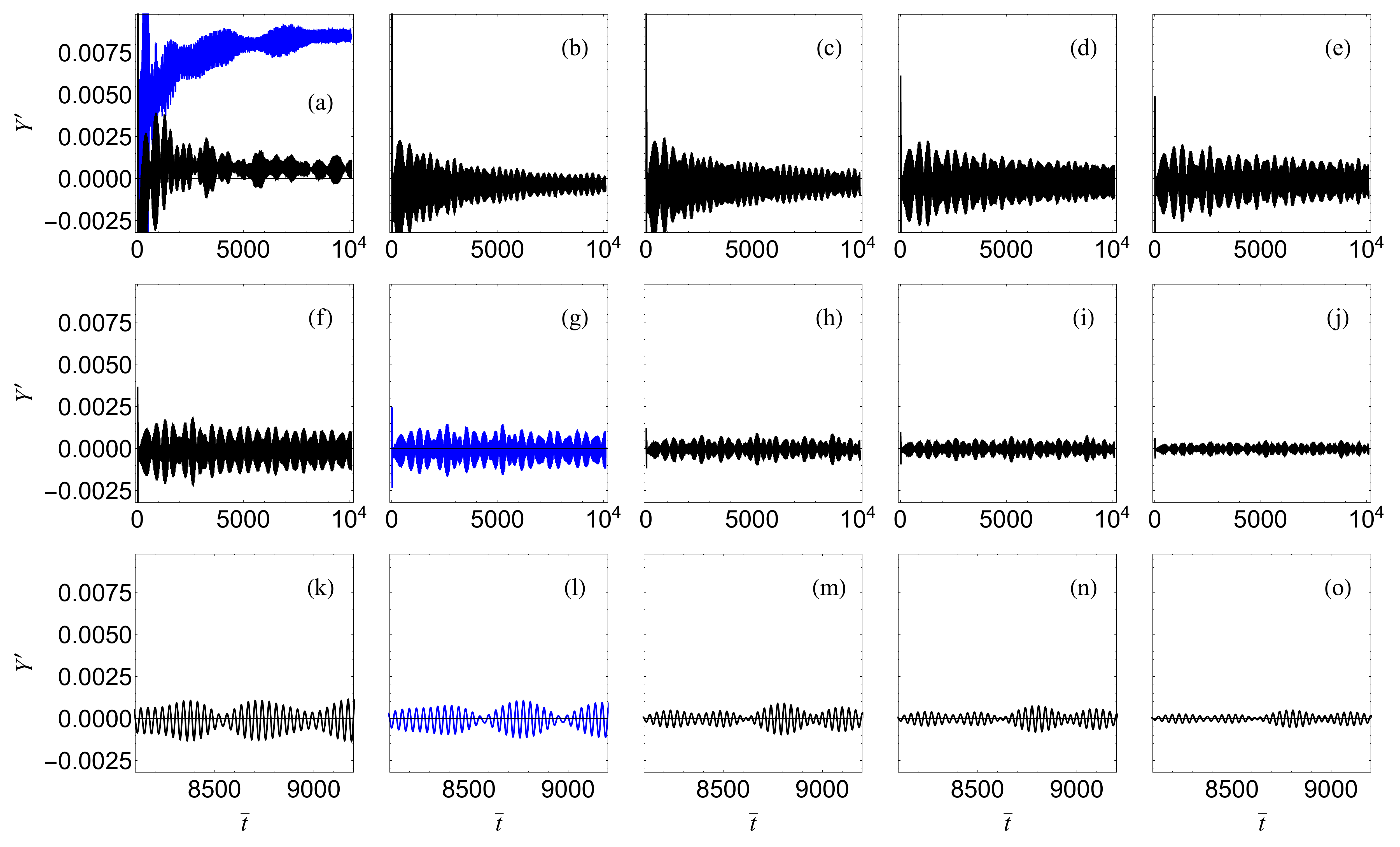}
\caption{Each plot is a time evolution of $Y' = \partial Y/\partial {\bar{r}}$, which is proportional to the radial electric charge density, for the specific value $\bar{r} = 5.01$ and the same initial data given in the caption of Fig.\ \ref{fig:evo 1}, except with the perturbation strength $f_2$ equal to (a) (from top to bottom) 1, 0.5 (b) 0.1, (c) 0.08, (d) 0.05, (e) 0.04, (f) 0.03, (g) 0.02, (h) 0.01, (i) 0.008, and (j) 0.005.  (a) is the same evolution shown in Fig.\ \ref{fig:evo 1} and (g) is the same evolution shown in Fig.\ \ref{fig:evo 2}.  The bottom row is the same as the middle row except zoomed in so that individual oscillations can be seen.}
\label{fig:w2 1}
\end{figure*}
\begin{figure*}
\centering
\includegraphics[width=6.5in]{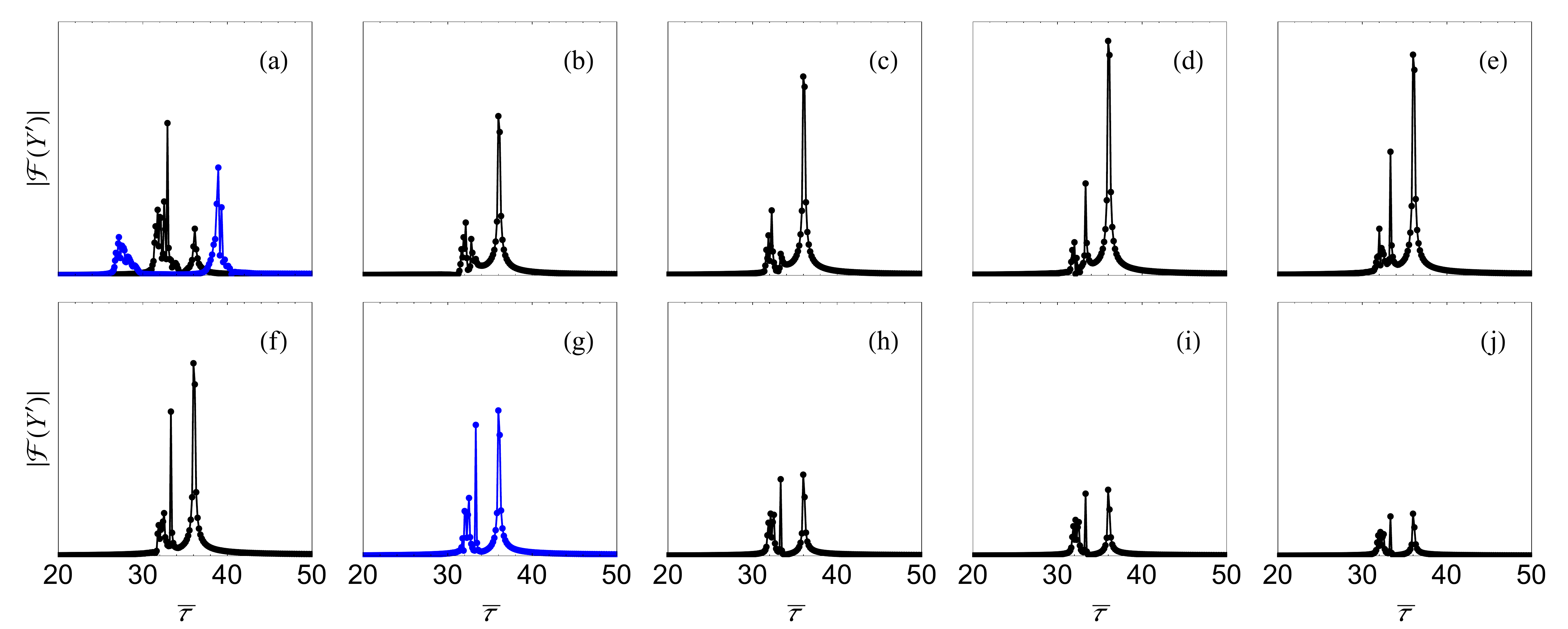}
\caption{Each plot gives the Fourier transform of the data (for $\bar{t} > 2000$) shown in the corresponding plot in the first two rows of Fig.\ \ref{fig:w2 1}.  (The vertical scale is arbitrary, but consistent across the plots.)  Once the perturbation size is sufficiently small, there are always two narrow spikes with periods $\bar{\tau}_1 =33.3 \pm 0.1$ and $\bar{\tau}_2 =36.0 \pm 0.1$, which are independent of the perturbation size.}
\label{fig:w2 FFT}
\end{figure*}
\begin{figure*}
\centering
\includegraphics[width=6.5in]{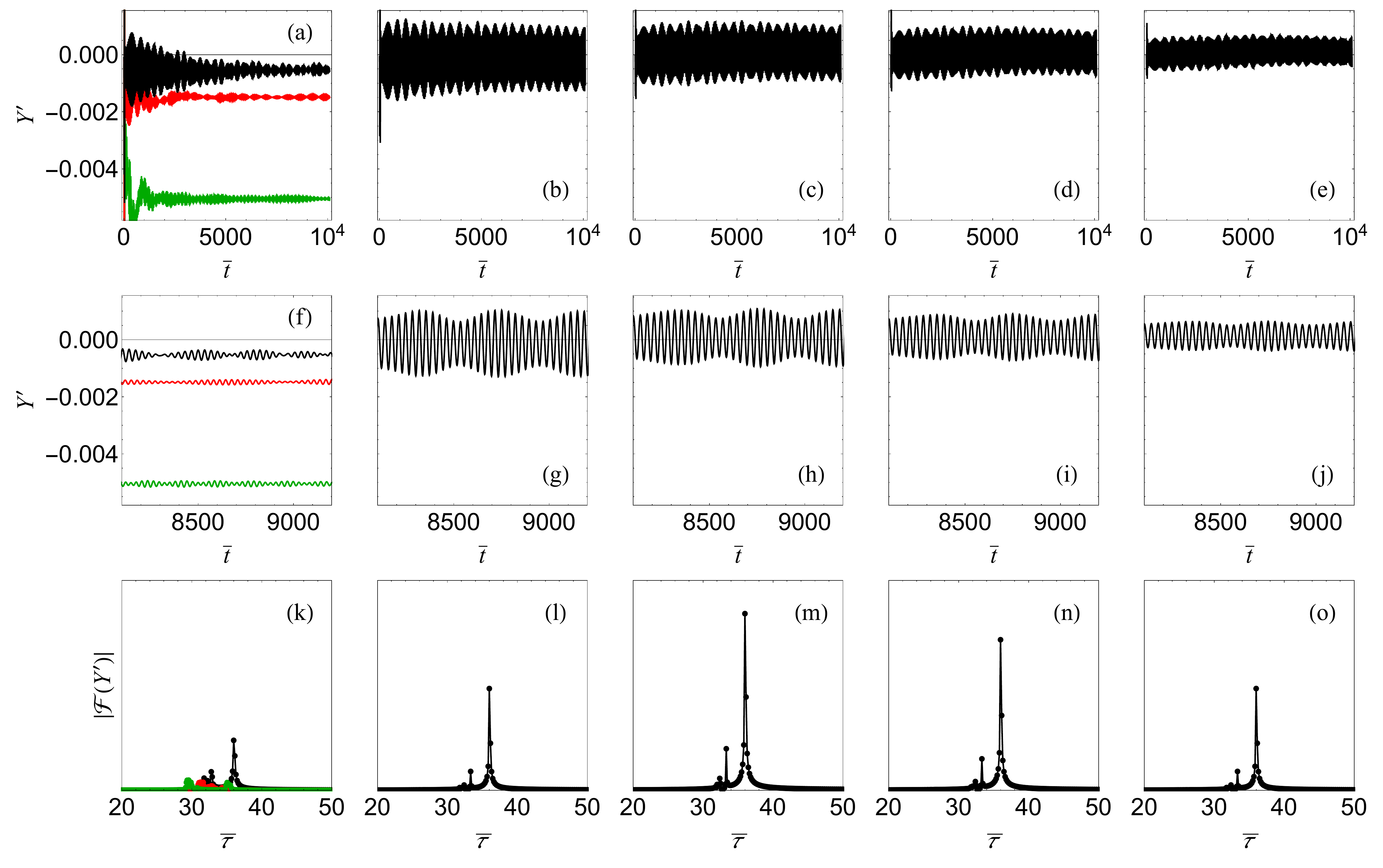}
\caption{Each plot in the top row is a time evolution of $Y' = \partial Y/\partial {\bar{r}}$, which is proportional to the radial electric charge density, for the specific value $\bar{r} = 5.01$.  The initial data for the evolutions is the regular static monopole solution with $\bar{v} = 0.2$, $\lambda = 0$, and sphaleronic perturbation (\ref{ur IC}) with $\bar{r}_r = 25$, $\bar{s}_r = 10$, and perturbation strengths $\bar{f}_r$ equal to (a) (from bottom to top) 1, 0.5, 0.3, (b) 0.2, (c) 0.1, (d) 0.08, and (e) 0.05.  The middle row is the same as the top row except zoomed in so that individual oscillations can be seen.  The bottom row is the Fourier transform of the top row for $\bar{t} > 2000$.  (The vertical scale for $|\mathcal{F}(Y')|$ is arbitrary, but consistent across the plots.)  As in Fig.\ \ref{fig:w2 FFT}, once the perturbation size is sufficiently small there are always two narrow spikes with periods $\bar{\tau}_1 =33.3 \pm 0.1$ and $\bar{\tau}_2 =36.0 \pm 0.1$, which are independent of the perturbation size.  }
\label{fig:ur}
\end{figure*}

As mentioned above, in analyzing stability with respect to sphaleronic perturbations I focus on $Y'$.  I show an alternative view of the evolutions of $Y'$ in Fig.\ \ref{fig:w2 1}.  The top curve in Fig.\ \ref{fig:w2 1}(a) is for the same evolution shown in Fig.\ \ref{fig:evo 1} and Fig.\ \ref{fig:w2 1}(g) is for the same evolution shown in Fig.\ \ref{fig:evo 2}.  The plots in Fig.\ \ref{fig:w2 1} are all for the specific value $\bar{r}=5.01$ and the first two rows show a series of evolutions with decreasing perturbation strengths.  One can see that as the size of the perturbation decreases, the oscillations of $Y'$ move toward being around zero.

Figure \ref{fig:evo 2} and the middle row of Fig.\ \ref{fig:w2 1}  are suggestive of oscillations about a stable equilibrium.  If this is the case, the stable equilibrium appears to be $Y' = 0$, which is the magnetic ansatz.  To further analyze this possibility I take a closer look at the individual oscillations.  The bottom row of Fig.\ \ref{fig:w2 1} shows the same evolutions as the middle row except zoomed in so that individual oscillations can be seen.  If these oscillations do actually contain harmonic oscillations about a stable equilibrium, we would expect a number of things about the period of the oscillations.  One would expect the period to be both $\bar{t}$ and $\bar{r}$ independent as well as independent of the strength of the perturbation (as long as the perturbation is sufficiently small).  A precise determination of the periods of the oscillations can be made from the Fourier transform, which I will label $\mathcal{F}(Y')$.  My interest is in the most rapid oscillations and I show in Fig.\ \ref{fig:w2 FFT} the Fourier transform of the data given in the first two rows of Fig.\ \ref{fig:w2 1}.  The Fourier transform presented is for data with $\bar{t} > 2000$, so as to ignore initial transient effects which precede the steady-state oscillations.  I find two narrow spikes with periods $\bar{\tau}_1 =33.3 \pm 0.1$ and $\bar{\tau}_2 =36.0 \pm 0.1$.  As the strength of the perturbation decreases, the locations of the spikes do not change and thus the periods of the oscillations are independent of the strength of the perturbation (as long as the perturbation is sufficiently small).  Further, I have Fourier transformed the data at different values of $\bar{r}$ and for different ranges of $\bar{t}$ and found the locations of the two spikes to be both $\bar{t}$ and $\bar{r}$ independent.  That there are two spikes whose periods are very close is expected after seeing beats in Fig.\ \ref{fig:w2 1}.

I now perturb the well-known static gravitating monopole solutions.  These solutions were first studied in \cite{VanNieuwenhuizen:1975tc, Lee:1991vy, Ortiz:1991eu, Breitenlohner:1991aa}, with a comprehensive analysis given in \cite{Breitenlohner:1991aa, Breitenlohner:1994di}.  I gave a limited review, using the same notation used here, of constructing the solutions in \cite{kain}.  In terms of matter fields, the static monopole solutions have nonzero values for $\varphi$, $\Phi$, $w_1$, and $Q_1$.  After constructing a static solution for the initial data, I add a $u_r$-sphaleronic perturbation, which is again a Gaussian:
\begin{equation} \label{ur IC}
u_r(0,r) = f_r 
(r/r_r )
e^{-(r-r_r)^2/s_r^2},
\end{equation}
along with $\dot{u}_r(0,r) = 0$.  The parameters $r_r$ and $s_r$ give the center and spread of the perturbation and $f_r$ can be thought of as its strength.  I note that this perturbation gives a nonzero value for $Q_2$, as can be seen from (\ref{1st order var def}).

I show the evolution of $Y'$ for a series of perturbations of decreasing size in Fig.\ \ref{fig:ur}, again for $\bar{r} = 5.01$.  In the top row we see that as the perturbation strength decreases the oscillations of $Y'$ move toward being around zero.  The middle row of Fig.\ \ref{fig:ur} presents the same evolutions as the top row except zoomed in so that individual oscillations can be seen.  The bottom row shows the Fourier transform of the top row for $\bar{t} > 2000$.  I find spikes in the Fourier transform at the exact same locations that we did in Fig.\ \ref{fig:w2 FFT}:\ $\bar{\tau}_1 = 33.3 \pm 0.1$ and $\bar{\tau}_2 = 36.0 \pm 0.1$.  I have Fourier transformed this data at different values of $\bar{r}$ and for different ranges of $\bar{t}$ and found the locations of the two spikes to be both $\bar{t}$ and $\bar{r}$ independent.

The results in this section are evidence for the stability of gravitating magnetic monopoles with respect to sphaleronic perturbations.  As the sphaleronic perturbation decreases in size, the system is found to oscillate about the magnetic ansatz.  For sufficiently small perturbations the periods of the oscillations are both $\bar{t}$ and $\bar{r}$ independent and independent of the strength of the perturbation.  Indeed, the periods are independent of the initial data entirely.  Two distinct and narrow spikes in the Fourier transform were found with periods $\bar{\tau}_1 = 33.3 \pm 0.1$ and $\bar{\tau}_2 = 36.0 \pm 0.1$.

In this section I displayed results only for $\bar{v} = 0.2$ and $\bar{\lambda} = 0$.  I have looked at other values of $\bar{v}$ (but kept $\bar{\lambda} = 0$) and found that the oscillation periods depend on $\bar{v}$, but are otherwise independent of initial data (for sufficiently small perturbations).  Indications are that there exists a branch of static monopole solutions (parametrized by $\bar{v}$) that are stable with respect to sphaleronic perturbations.  Mapping this out is beyond the scope of this work, but it would be interesting to look at this more closely.


\section{Conclusion}
\label{sec:conclusion}

$SU(2)$ with a scalar field in the adjoint representation coupled to gravity has as a classical solution the gravitating Julia-Zee dyon \cite{Julia:1975ff, Brihaye:1998cm,Brihaye:1999nn}.  Within the magnetic ansatz this system no longer contains dyons and instead has as a classical solution the gravitating 't Hooft-Polyakov magnetic monopole \cite{tHooft:1974kcl,Polyakov:1974ek, VanNieuwenhuizen:1975tc, Lee:1991vy, Ortiz:1991eu, Breitenlohner:1991aa, Breitenlohner:1994di}.  I developed second-order code to dynamically solve the full gravitating dyon system.  As far as I am aware, this is the fist time the gravitating dyon has been dynamically solved and is one of the only times (aside from the important papers of Rinne et al.\ \cite{Rinne:2013qc, Maliborski:2017jyf}) that $SU(2)$ has been dynamically solved outside the magnetic ansatz.

In pure $SU(2)$ (i.e.\ unbroken and without the scalar field), regular static gravitational solutions are known as Bartnik-McKinnon solutions \cite{Bartnik:1988am}, which are well-known to be unstable with respect to both magnetic \cite{Straumann:1989tf} and sphaleronic \cite{Volkov:1994dq, Lavrelashvili:1994rp, Volkov:1995np} perturbations.  The stability of regular static gravitating monopoles with respect to magnetic perturbations was studied by Hollmann \cite{Hollmann:1994fm}, who found that they are always stable for values of $\bar{v}$ not too large.  Other studies corroborated this \cite{Maeda:1993ap, Tachizawa:1994wn, kain}, leaving little question as to their stability with respect to magnetic perturbations.   As far as I am aware, the stability of gravitating monopoles with respect to sphaleronic perturbations has not been studied and thus it is an open question whether gravitating monopoles are stable.  That it has not been studied is presumably because it is very challenging to do so.  Indeed, \cite{Aichelburg:1992st, Maeda:1993ap, Tachizawa:1994wn, Hollmann:1994fm} indicated that a standard harmonic stability analysis of magnetic perturbations (let alone sphaleronic perturbations) is very difficult to perform (semi)analytically.

I chose to avoid these difficulties by making a purely numerical study of sphaleronic stability.  I did this by adding a sphaleronic perturbation to both generic magnetic initial data and the static gravitating monopole solutions \cite{VanNieuwenhuizen:1975tc, Lee:1991vy, Ortiz:1991eu, Breitenlohner:1991aa, Breitenlohner:1994di} and then evolving the system.  For large perturbations the system heads away from the gravitating monopole and toward a dyon configuration.  As the perturbation decreases in size, the system begins oscillating about the magnetic ansatz.  I found that the periods of the oscillations are independent of time, position, and initial data (as long as the perturbation is sufficiently small), exactly what one would expect for oscillations about a stable equilibrium.  I thus found numerical evidence for gravitating monopoles being stable with respect to sphaleronic perturbations.  

A numerical stability analysis can rarely replace an analytical one and the results presented here do not prove that gravitating monopoles are stable.  Nevertheless, the results are, as far as I am aware, the first piece of evidence discovered for the possible stability of gravitating monopoles with respect to sphaleronic perturbations. 

In this work I did not consider black holes.  There exist static gravitating black hole monopole solutions \cite{Lee:1991vy, Breitenlohner:1991aa, Breitenlohner:1994di} and it is an important question whether they too are stable with respect to sphaleronic perturbations.  There are a few reasons why I did not consider them.  Some of the reasons are numerical:\  the code I use is less stable when a black hole forms and it looks to be very difficult to construct initial data that is a perturbation of a static black hole monopole solution.  It is not difficult to construct initial data that is a sphaleronic perturbation of generic regular magnetic data which forms a black hole during the evolution.  However, the static black hole monopole solutions are not unique (there exists a continuum of solutions with the same $\bar{v}$ and $\bar{\lambda}$ \cite{Breitenlohner:1991aa}) and it is therefore not clear which solution is being perturbed in a given evolution.




%


\end{document}